%% file: Paper.tex
\documentclass[aps,prx,letterpaper,twocolumn,nofootinbib,showkeys,floatfix,amsmath,amssymb]{revtex4}

\usepackage{graphicx}
  \graphicspath{{./Figures/}}
\usepackage{hyperref}
  \hypersetup{colorlinks=true,citecolor=blue,linkcolor=blue,urlcolor=blue}

\usepackage{cleveref}
\usepackage{mathtools}
\usepackage{physics}

\crefname{step}{step}{steps}
\newtheorem{theorem}{Theorem}
\DeclarePairedDelimiter{\ceil}{\lceil}{\rceil}
\DeclarePairedDelimiter{\floor}{\lfloor}{\rfloor}
\newcommand{\defeq}{\triangleq}
\newcommand{\ra}{\rightarrow}
\newcommand{\LRa}{\Leftrightarrow}
\newcommand{\qra}{\quad \rightarrow \quad}
\newcommand{\qRa}{\quad \Rightarrow \quad}

\begin{document}

\preprint{IIT-CAPP-20-06}

\title{An optimal quantum sampling regression algorithm for variational eigensolving\\ in the low qubit number regime}

\author{Pedro Rivero}
\email{priveroramirez@anl.gov}
\affiliation{Department of Physics, Illinois Institute of Technology, Chicago, Illinois 60616-3793, USA}
\affiliation{Physics Division, Argonne National Laboratory, Lemont, Illinois 60439, USA}

\author{Ian C.\ Clo\"{e}t}
\affiliation{Physics Division, Argonne National Laboratory, Lemont, Illinois 60439, USA}

\author{Zack Sullivan}
\affiliation{Department of Physics, Illinois Institute of Technology, Chicago, Illinois 60616-3793, USA}

\date{December 3, 2020}

\begin{abstract}
  The VQE algorithm has turned out to be quite expensive to run given the way we currently access quantum processors (i.e. over the cloud). In order to alleviate this issue, we introduce Quantum Sampling Regression (QSR), an alternative hybrid quantum-classical algorithm, and analyze some of its use cases based on time complexity in the low qubit number regime. In exchange for some extra classical resources, this novel strategy is proved to be optimal in terms of the number of samples it requires from the quantum processor. We develop a simple analytical model to evaluate when this algorithm is more efficient than VQE, and, from the same theoretical considerations, establish a threshold above which quantum advantage can occur. Finally, we demonstrate the efficacy of our algorithm for a benchmark problem.
\end{abstract}


\maketitle

\input{Content/01_Introduction}
\input{Content/02_Algorithm-outline}
\input{Content/03_Complexity-analysis}
\input{Content/04_Low-qubit-number-regime}
\input{Content/05_Applications}
\input{Content/06_Benchmarking}
\input{Content/07_Conclusions}

\begin{acknowledgments}
  This work was partially supported by the Quantum Information Science and Engineering Network (\href{https://qisenet.uchicago.edu/overview/}{QISE-NET}); which is co-led by the University of Chicago and Harvard University, managed by the Chicago Quantum Exchange, and funded by the National Science Foundation (NSF) as part of the agency's Quantum Leap Big Idea. P.R.\ and I.C.\ were supported by the U.S.\ Department of Energy, Office of Science, Office of Nuclear Physics, contract no.\ DE-AC02-06CH11357.
\end{acknowledgments}

\bibliography{PR}

\end{document}

%% file: Content/01_Introduction.tex

\section{Introduction}\label{sec:introduction}

One of the most promising near-term applications of quantum computers is the simulation of quantum mechanics. This idea of using quantum systems to simulate other such entities is at the root of Quantum Computing itself, and traces back to Richard Feynman's famous proposal of the Quantum Computer \cite{FEYNMAN:1982:simulating-physics-with-computers}.

The interest on this particular use stems from the fact that quantum phenomena are exponentially hard to recreate by classical means, whilst thought to be efficiently reproducible using quantum resources. This presents two complementary views of the problem. From a physics standpoint, we expect to be able to simulate nature at a level of detail unimaginable before, with powerful consequences for fields like quantum chemistry, or materials engineering. From the computational side of things, this task might be one of the first to prove that quantum computers are indeed more efficient than classical machines. That would mean reaching a long sought and highly desirable scientific landmark known as \emph{quantum advantage} \cite{PRESKILL:2011:quantum-computing-entanglement-frontier}. More importantly, said advantage would be reached for something useful beyond just showing the computational power of these devices.

The reason why physicists believe that this task can be accomplished sooner rather than later lays, nearly in its entirety, on the so called \emph{hybrid quantum-classical variational algorithms} \cite{PERUZZO.MCCLEAN.EA:2014:variational-eigenvalue-solver-photonic, FARHI.GOLDSTONE.EA:2014:quantum-approximate-optimization-algorithm, VERDON.BROUGHTON.EA:2019:quantum-algorithm-train-neural, BRAVO-PRIETO.LAROSE.EA:2020:variational-quantum-linear-solver}, whose notoriety is due to their promising applicability --- especially for relatively low amounts of quantum resources (i.e. near term and state of the art machines). This scenario where we have access to quantum processors, but only for small efforts due to their noisy nature, has come to be known in the literature as the \emph{Noisy Intermediate-Scale Quantum} era (NISQ) \cite{PRESKILL:2018:quantum-computing-nisq-era}. During this time, it is expected that quantum technologies will surpass classical ones. Nonetheless, NISQ technology will only show such advantage for a limited number of tasks before the threshold for performing fully fault-tolerant quantum computations, as predicted by the quantum fault-tolerance theorem \cite{AHARONOV.BEN-OR:1996:fault-tolerant-quantum-computation}, is overcome.

The theoretical insight upon which these algorithms have been built is the variational theorem of quantum mechanics \cite{PERUZZO.MCCLEAN.EA:2014:variational-eigenvalue-solver-photonic,MCCLEAN.ROMERO.EA:2016:theory-variational-hybrid-quantum-classical}:

\begin{theorem}\label{th:quantum-variational-theorem}

  If the state $\ket{\psi}$ of a quantum system depends on some array of $n$ parameters $\qty{\theta^{n}}$, the optimal choice to approximate the ground state of said system (i.e. the eigenstate of its Hamiltonian $\hat{H}$ with minimum eigenvalue $\lambda_{\textnormal{min}}$), is the one which minimizes its Hamiltonian's expectation value $\ev*{\hat{H}}$. Assuming $\braket*{\psi}=1$:

  \begin{gather}
    \ev{\hat{H}}\!\qty(\theta^{n}) \equiv
      \ev{\hat{H}}{\psi\qty(\theta^{n})} \geq
      \lambda_{\textnormal{min}} \, .
  \end{gather}

\end{theorem}

The same logic applies to finding the eigenstate with maximum eigenvalue: only maximizing the expectation value function instead. Furthermore, by transforming the target operator $\hat{H} \ra (\hat{H}-\gamma{\hat{1}})^2$ one can find any of its original eigenstates for different values of $\gamma$. There is, of course, nothing special about the Hamiltonian in an abstract mathematical way, and therefore this technique can be straightforwardly generalized to any Hermitian operator (e.g. quantum observables).


\subsection{Variational Quantum Eigensolver}\label{sub:vqe-algorithm}

\begin{figure}[!tbp]
  \centering
  \includegraphics[width=\linewidth]{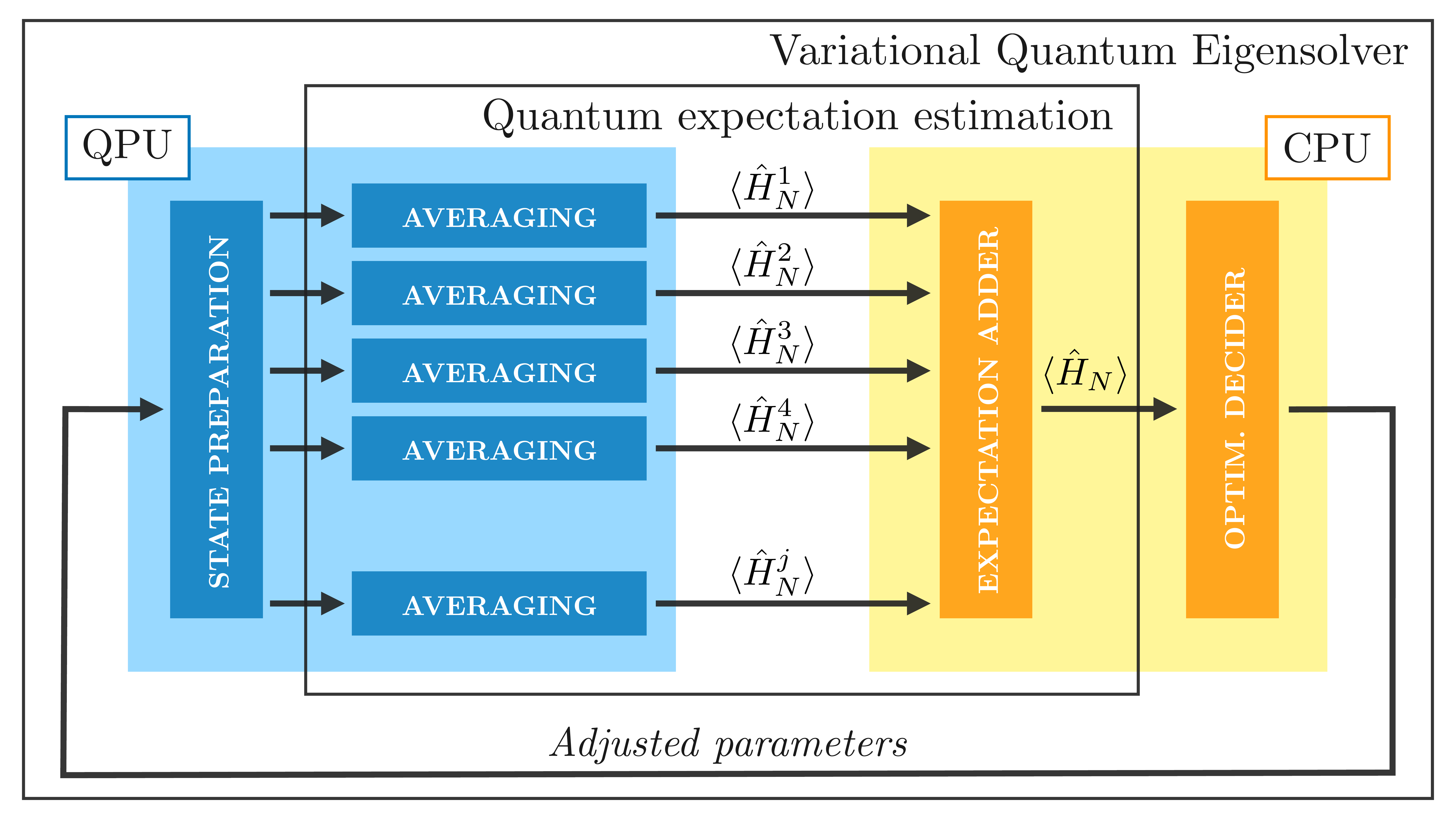}
  \caption{Diagrammatic representation of the Variational Quantum Eigensolver algorithm (VQE) as depicted in its original paper \cite{PERUZZO.MCCLEAN.EA:2014:variational-eigenvalue-solver-photonic}.}
  \label{fig:vqe-algorithm}
\end{figure}

The \emph{Variational Quantum Eigensolver} (VQE) \cite{PERUZZO.MCCLEAN.EA:2014:variational-eigenvalue-solver-photonic}, is probably the most well-known hybrid quantum-classical variational algorithm. It essentially works by taking advantage of quantum processors for performing some of the costly operations embedded within classical optimization schemes \cite{MCCLEAN.ROMERO.EA:2016:theory-variational-hybrid-quantum-classical} --- such as evaluating the objective function. While typically applied to solving quantum problems, it can also be used for other purposes by efficiently encoding any target question in a quantum-like formalism; in which case it goes by different names. When employed for classical combinatorial optimization it is commonly referred to as the \emph{Quantum Approximate Optimization Algorithm} (QAOA) \cite{FARHI.GOLDSTONE.EA:2014:quantum-approximate-optimization-algorithm}. All in all, the general outline of this algorithm is illustrated in \cref{fig:vqe-algorithm}, and works as follows:
\begin{enumerate}
  \item Prepare a quantum state in the quantum processor according to some parameters (e.g. an initial set of parameters).\label[step]{it:vqe-state-preparation}
  \item Evaluate the expectation value of the different computable addends making up the target operator (\cref{sub:operator-averaging}) by performing repeated measurements on the quantum processor.\label[step]{it:vqe-expectation-values}
  \item Combine the previous expectation values by adding them up via classical means according to their respective weights in the expression for the target operator. This will result in the expectation value of said observable for the prepared state; see \cref{eq:vqe-operator-averaging}.\label[step]{it:vqe-weighted-combination}
  \item Use a classical optimization decider to analyze this objective function (i.e. the expectation value of the target operator for different states) and generate a new set of parameters ``closer'' to those for which the objective function evaluates to a global minimum.\label[step]{it:vqe-classical-optimization}
  \item Return to step one or stop if convergence has been reached. The desired eigenstate of the system is described by the set of parameters returned by the classical optimizer once the algorithm has converged, along with the state preparation strategy implemented; while the eigenvalue corresponds to the expectation value of the target operator for such normalized state.\label[step]{it:vqe-convergence-iteration}
\end{enumerate}

In practice, depending on the classical optimization scheme that we choose (e.g. COBYLA, steepest descent, Nelder-Mead simplex, BFGS, genetic \cite{NOCEDAL.WRIGHT:2006:numerical-optimization,MITCHELL:1998:introduction-genetic-algorithms}), the actions in \cref{it:vqe-state-preparation,it:vqe-expectation-values,it:vqe-weighted-combination} will be repeated several times in order to perform the task in \cref{it:vqe-classical-optimization}; however the general layout of the algorithm will remain unchanged. This is the reason why using quantum resources for these undertakings has such a powerful impact on the overall efficiency of the optimization. Particularly, the classical evaluation of the expectation value of a discrete quantum observable $\hat{H}_{N}$ for a given state $\ket{\psi}$, takes time which grows exponentially with the size of system (i.e. the number of qubits $N$ used to represent it); while often being significantly more efficient on a quantum processor (e.g. polynomial time).


\subsection{Operator averaging}\label{sub:operator-averaging}

Evaluating the expectation value of the target operator needs to be done efficiently. Generally speaking, quantum computers cannot measure any given observable directly, and so it becomes necessary to re-express ours in terms of other operators which can indeed be probed on such devices --- typically Pauli operators. From there, the task is carried out through a process known as \emph{operator averaging}; which, as the name suggests, consists on taking a high number of measurements (referred to as ``shots'') and averaging the results in the usual fashion. On top of that, general statistical analysis can be performed in order to predict the convergence and statistical error of the recovered results \cite{MCCLEAN.ROMERO.EA:2016:theory-variational-hybrid-quantum-classical}.

By looking at the action described in \cref{it:vqe-weighted-combination} we can see why it is important that the number of addends of the refactored operator grows at most polynomially with the size of the system. If this was not the case, said step could not be performed efficiently, and we would not be able to simulate our system effectively on growing values of the number of qubits used to represent it. Altogether, we can synthesize this by saying that, for $q \in {\mathbb{R}}^{+}$, our target operator must obey the following:

\begin{gather} \label{eq:vqe-operator-averaging}
  \hat{H}_{N} = \sum_{j=1}^{\order{N^q}} w_{j} \hat{H}_{N}^{j} \qRa
  \ev{\hat{H}_{N}} = \sum_{j=1}^{\order{N^q}} w_{j} \ev{\hat{H}_{N}^{j}} \, .
\end{gather}

Where $\hat{H}_{N}^{j}$ are $N$-qubit operators which can be directly measured on a specific quantum computer. Note that we are making use of the correspondence between polynomial complexity and asymptotic efficiency \cite{KLEINBERG.TARDOS:2006:algorithm-design}.


\subsection{Space parametrization and state preparation}\label{sub:space-parametrization-and-state-preparation}

Even if the expectation value of an operator for a given state can now be efficiently evaluated, many different such states may still be needed during the optimization process. This is why the parametrization required for \cref{it:vqe-state-preparation} plays a critical role in the algorithm. Since the dimensionality of phase space grows exponentially with the number of qubits (i.e. $n=\order*{2^N}$), we will be interested in finding a smaller subspace containing the solution to our problem. This makes the task of parametrizing the different quantum states far from trivial, often relying on heuristics and theoretical ansatzes to narrow down the searchable region of phase space. In many cases, taking advantage of the symmetries of the target operator has been proven very fruitful although not general. In fact, extensive efforts are taking place at present in order to improve current procedures \cite{KANDALA.MEZZACAPO.EA:2017:hardware-efficient-variational-quantum-eigensolver, SETIA.CHEN.EA:2019:reducing-qubit-requirements-quantum}.

Beyond simply exploring the interesting states of our system, notice that the dependence of the objective function $\ev*{\hat{H}_{N}}$ on these parameters will greatly influence the convergence of any optimization scheme used. A good choice of parametrization ansatz will make the resulting expectation value function ``easy'' to optimize, avoiding unnecessary local minima and other hazards. Furthermore, we can exploit this relation in order to predict the general form that the function will assume under certain parametrization; taking advantage of that knowledge then for developing a new hybrid quantum-classical variational algorithm. We tackle said goal in the present work.

Ultimately, the whole process relies on our ability to efficiently prepare quantum states according to these ansatz parameters; a task that is typically engineered alongside space parametrization, and carried out using quantum circuits. For this reason, the complexity of said circuits is often required to be polynomial as well. The interested reader can find more valuable information on this topic in \cite{MCCLEAN.ROMERO.EA:2016:theory-variational-hybrid-quantum-classical, LEE.HUGGINS.EA:2019:generalized-unitary-coupled-cluster, ASPURU-GUZIK.DUTOI.EA:2005:simulated-quantum-computation-molecular, ROMERO.BABBUSH.EA:2017:strategies-quantum-computing-molecular, HELGAKER.JORGENSEN.EA:2000:molecular-electronic-structure-theory}.


\subsection{Motivation and thesis}\label{sub:motivation-and-thesis}

One of the biggest practical setbacks of VQE comes from the way we currently access quantum processors (i.e. through the cloud). The need to establish an internet connection on every cycle of the optimization loop makes execution unnecessarily time consuming; especially if users are pushed into a queue for every required sample of the objective function. Some backend providers, such as IBM, are already working on functionalities that modify optimization processes so that all related samples within the same cycle can be requested in a single query (e.g. to obtain numerical derivatives). This is progress, however the approach is laborious to extend to optimization methods beyond those provided by default. Even worse, it does not address the most significant part of this bottleneck, which is intrinsic to the algorithm itself. Due to the way in which VQE is structured, we cannot know what the next sample (or set of samples) is going to be before asking for the current one.

Our goal with this work is to develop an alternative algorithm that gets rid of such a bottleneck, even if more classical resources are needed. Given that access to quantum computers is still a low-offer/high-demand commodity, we deem this trade-off both possible and convenient for many current studies --- harnessing all available classical capabilities, while still capturing many of the inner workings and intrinsic properties of quantum computations that do not appear in simulation. Additionally, this better tuned balance between quantum and classical capacities, will shine some light into the question of quantum advantage and when it can be achieved.

%% file: Content/02_Algorithm-outline.tex

\section{Algorithm outline}\label{sec:algorithm-outline}

As we mentioned in \cref{sub:space-parametrization-and-state-preparation}, if we are smart when parametrizing phase space, we will be able to foresee the general form of the expectation value function that we are trying to optimize. Particularly, from the use rotations in our quantum circuit one can predict cycles in the the states that we are parametrizing. Such periodic nature on a bounded domain (i.e. $\theta^{n} \! \in \; ]-\pi, \pi]^{\otimes n}$) will transfer to the expectation value function, which in turn allows us to consistently apply Fourier analysis on a discrete amount of frequencies to fully describe it. On top of that, by looking at how the different rotations interact with one another in the quantum circuit enforcing our parametrization (i.e. the circuit topology), we can anticipate the maximum frequency that the resulting function will present. Typically, this will be some integer multiple of the natural/fundamental angular frequency (i.e. first harmonic), which for a regular rotation simply happens to be $\omega_{1} = 1$. We will later see examples of this to illustrate our point.

Having a bounded bandwidth and a finite amount of frequencies suggests implementing a linear regression over sinusoids (i.e. Fourier basis) to recover the complete form of the function as a trigonometric series. This strategy makes use of our knowledge about the topology of the domain to reduce the number of states that need to be sampled on the quantum processor. Furthermore, in some cases, once the superposition function approximating the objective distribution is found, it will be easier to infer its global minimum; either through theoretical considerations, heuristics, or a mix of these two. We will develop this formalism for only one ansatz parameter first, as that will suffice to make our point clear, and the generalization to higher dimensions is straight forward keeping in mind a few considerations that we will address afterwards. With all this in mind, we can write:

\begin{gather}
  h(\theta) \defeq \ev{\hat{H}_{N}}\qty(\theta) \equiv a_0 + \sum_{k=1}^S \qty[a_k\cos(k\theta) + b_k\sin(k\theta)] \, .
\end{gather}

Generally $S \ra \infty$, however, if the bandwidth is bounded, $S$ will be finite and it will be possible to evaluate this expression exactly. We will assume this to be the case for the time being and return to it later. We can now sample the objective function $h(\theta)$ several times and solve for the Fourier coefficients $\qty{a_0,a_k,b_k} \forall k$. Particularly, we will need $2S+1$ samples in order to have as many equations as unknowns. Also, we would like to choose those samples as to mitigate any sampling errors, especially around the global minima. To achieve this we can follow two approaches other than improving the quality of the samples themselves (e.g. incrementing the number quantum measurements or shots used to compute the expectation values): increase the total amount of samples, or optimize the spacing between them (i.e. the lattice). All in all, this will result in a linear system of the form shown in \cref{eq:regression-linear-system}.

\begin{widetext}
  \begin{gather}\label{eq:regression-linear-system}
    \mqty[
      1 & \cos(\theta_1) & \sin(\theta_1) & \cos(2\theta_1)
        & \cdots & \sin(S\theta_1) \\
      1 & \cos(\theta_2) & \sin(\theta_2) & \cos(2\theta_2)
        & \cdots & \sin(S\theta_2) \\
      \vdots & \vdots & \vdots & \vdots & \ddots & \vdots \\
      1 & \cos(\theta_{2S+1}) & \sin(\theta_{2S+1}) & \cos(2\theta_{2S+1})
        & \cdots & \sin(S\theta_{2S+1})
    ]
    \mqty[
      a_0 \\ a_1 \\ b_1 \\ a_2 \\ \vdots \\ b_S
    ] =
    \mqty[
      h(\theta_1) \\ h(\theta_2) \\ \vdots \\ h(\theta_{2S+1})
    ]
  \end{gather}
\end{widetext}

Although this system should be solvable if the assumptions were correct, in practice, because our sampling process is flawed (i.e. we have neither infinite precision nor time to calculate the exact expectation values), it will most likely turn out inconsistent. For this reason, it is a good idea to normalize these equations before solving; which is equivalent to finding the \emph{least-squares solution} of the linear system \cite{BURGOS:2006:algebra-lineal-geometria-cartesiana}. The idea behind normalizing the equations is to project (orthogonally) the non-homogeneous vector over the linear mapping's range. The resulting equations will then be solvable but they may have several solutions. If that was the case further processing would need to be carried out to get a so called \emph{optimal solution} of the linear system (i.e. a solution orthogonal to the kernel of the linear mapping).

\begin{gather}
  Fc = h \qra F^{\dagger}Fc = F^{\dagger}h \, .
\end{gather}

Extending these results to higher dimensions is straightforward considering multidimensional Fourier series. In this case, we may have a different bandwidth $S_{j}$ for each ansatz parameter. For a total number of ansatz parameters $n$ (i.e. dimensions), and a maximum bandwidth $S_{\text{max}}$, the total number of samples required to perform the linear regression is:

\begin{gather} \label{eq:QSR-algorithm-samples}
  T = \prod_{j=1}^{n} \qty(2 S_{j} + 1) \leq
    \qty(2 S_\textnormal{max} + 1)^{n} \equiv 2^{sn} \, .
\end{gather}

Where we have defined $s$ as a non-strict upper bound to the necessary number of samples, expressed in bits, per dimension. For instance:

\begin{gather} \label{eq:QSR-sample-parameter}
  s \equiv \log_2 \qty(2S_\textnormal{max} + 1) \geq \frac{\log_2 T}{n} \, .
\end{gather}

\begin{figure}[!btp]
  \centering
  \includegraphics[width=\linewidth]{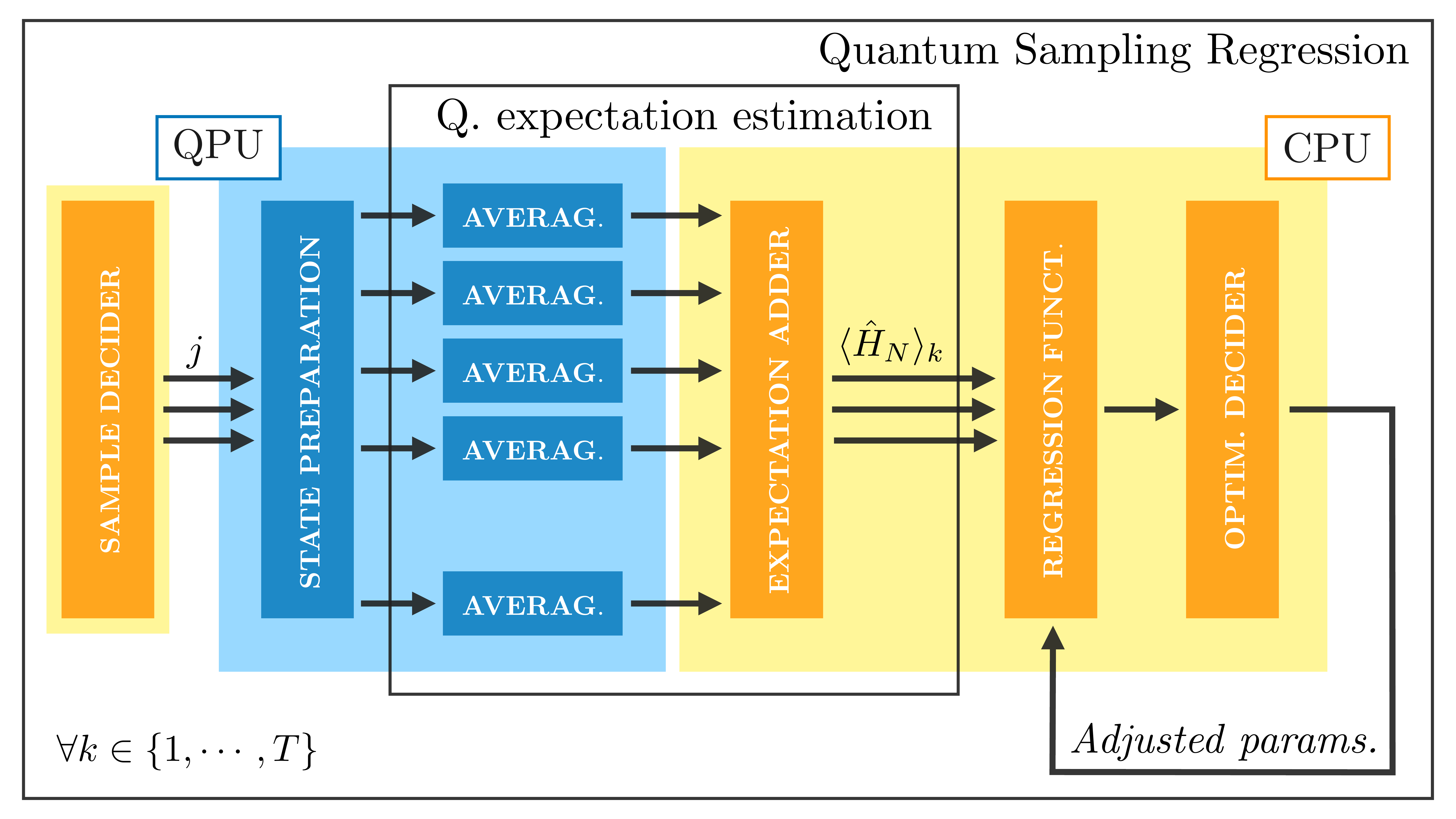}
  \caption{Diagrammatic representation of the Quantum Sampling Regression algorithm (QSR).}
  \label{fig:QSR-algorithm}
\end{figure}

We can now harness this method to construct a new hybrid quantum-classical variational eigensolver which we shall refer to as the \emph{Quantum Sampling Regression} algorithm (QSR). Its general layout, as illustrated in \cref{fig:QSR-algorithm}, is:
\begin{enumerate}
  \item Determine the bandwidth associated to each parameter in the ansatz; an upper bound would suffice.
  \item Sample the objective function at least as many times as shown in \cref{eq:QSR-algorithm-samples} using a quantum processor in the same way as for VQE (e.g. evenly spaced samples).
  \item Compute the Fourier coefficients from the samples (i.e. solve the normalized linear system of equations).
  \item In a classical machine, solve for the global minimum of the resulting regression function.
\end{enumerate}

Theoretically, the power of this algorithm, as opposed to impromptu exhaustive search, is demonstrated through the \emph{Nyquist-Shannon sampling theorem} \cite{SHANNON:1949:communication-presence-noise}. This method is equivalent to ours in terms of the number of sample points needed (i.e. one more than the strict lower bound $2S$); which proves the optimality of our strategy for fully reconstructing the desired objective function.

\begin{theorem}\label{th:nyquist-shannon-sampling-theorem}

  If a function $h(\theta)$ contains no angular frequencies higher than $\omega_{\textnormal{S}}$, it is completely determined by giving its ordinates at a series of points $1/2\omega_{\textnormal{S}}$ apart:

  \begin{gather}
    \omega_{\textnormal{sampling}} > 2\omega_{\textnormal{S}} \, .
  \end{gather}

\end{theorem}

%% file: Content/03_Complexity-analysis.tex

\section{Complexity analysis}\label{sec:complexity-analysis}

In order to assess the potential usefulness of this technique, it is important to contrast its efficiency with that of regular VQE, as well as classical computation. To do so, in this section, we will look at the asymptotic time complexity of these three different strategies; making some remarks about space complexity along the way if necessary. The decision to focus on time complexity is justified by the nature of these hybrid quantum-classical algorithms: as they do not store any intermediate quantum information in classical memory, and there are no other serious limitations given the amount of retrievable information handled by state of the art quantum technologies (i.e. while computation requires exponentially increasing classical memory, only a small fraction of it can actually be retrieved \cite{HOLEVO:1973:bounds-quantity-information-transmitted}). In the case of purely classical computation we will assume unlimited memory, since otherwise we would be facing a discussion on quantum advantage. Nonetheless, the reader must keep in mind that classical computers might, at some point, reach an operational memory limit which makes them incapable of keeping up with quantum computers regardless of the time complexity they present \cite{PRESKILL:2018:quantum-computing-nisq-era}.


\subsection{Classical computation}\label{sub:classical-complexity}

Although more elaborate methods may exists for dealing with quantum systems in classical machines, for the sake of simplicity, we will make a general argument about working with the usual matrix formulation of quantum mechanics.

The size of vectors representing quantum states of a system, as well as that of matrices representing observables, grows exponentially with the number of qubits $N$ (i.e. space complexity $\omega(2^{N})$). Since performing matrix-vector multiplication is an effort with quadratic time complexity on the size of said objects, we can introduce a lower bound for eigenvalue finding: $\Omega((2^{N})^2) = \Omega(4^{N})$. On the other hand, general methods for spectral decomposition of a given matrix, such as \emph{Householder transformations} or \emph{Givens rotations}, have cubic time complexity \cite{ARBENZ:2012:lecture-notes-solving-large, PRESS.TEUKOLSKY.EA:2007:numerical-recipes-c-art}. Accordingly, an upper bound would be: $\order*{(2^{N})^3} = \order*{8^{N}}$. We can see here how classical computation shows the usual exponential time complexity from emulating quantum systems, as well as exponential space complexity.

Notice that, in this approach, we do not need to parametrize Hilbert space, and therefore the results are irrespective of any number of parameters. Nonetheless, we could, in theory, simulate the parametrized optimization process from VQE using the matrix formalism for all evaluations of the expectation value function. This would not be an efficient approach though, since, in general, we would be nesting an exponentially hard process (i.e. matrix multiplication) inside the optimization loop: the former depending on the number of qubits $N$, and the latter on the number of ansatz parameters $n$. Even if the optimization process turned out to converge in polynomial time, that would still leave us with an exponential time algorithm: $\order*{4^{N}n^{p}}$. Where we are yet to find a relation between the number of ansatz parameters and the number of qubits $n(N)$. We will later address how QSR can help in fully exploiting the parametrization ansatz on classical machines, without resorting to this burdensome matrix formulation.

Altogether, there is no way to make this algorithm more efficient than exponential irrespective of $n(N)$. As a matter of fact, the most general case would be equivalent to dimension counting (i.e. $n=\order*{2^{N}}$), which leads, for $p=1$, to the same time complexity as before: $\order*{8^{N}}$; and even worse if the optimization was superlineal (i.e. $\omega(N)$; e.g. $p>1$). Notice that we are not simulating the quantum circuit, assuming the parametrization to be directly implemented in memory over the vector representing the state of the system. In a similar fashion, we are evaluating the target operator directly, instead of following the decomposition shown in \cref{eq:vqe-operator-averaging}.


\subsection{Variational Quantum Eigensolver}\label{sub:vqe-complexity}

Since the complexity of classical optimization processes might vary significantly from one optimizer to another, and from one problem to another, it is good to focus first on the task of evaluating the target operator (i.e. the unit steps that go inside the optimization loop). The reader must keep in mind though, that this would not be taking into account the usually most expensive part of the algorithm; which we will later have to reintroduce.

As a matter of fact, the most general instances of mathematical optimization are considered to be NP-hard \cite{MURTY.KABADI:1987:some-np-complete-problems-quadratic}; often taking exponential time to complete (e.g. exhaustive search) --- a phenomenon usually referred to as the \emph{curse of dimensionality} \cite{BELLMAN:1957:dynamic-programming, LIU.ZSU:2009:encyclopedia-database-systems}. This makes for the single major skepticism surrounding the VQE; concerns being even more severe for QAOA, where one does not even have the requirement of evaluating a quantum system in each step to begin with. Nevertheless, this also embodies the appeal of VQE: since using a quantum processor to probe a quantum system could, in principle and on its own, introduce savings unsurmountable by any classical computation if the optimization process was efficient.

As explained in \cref{sub:operator-averaging}, evaluating the expectation value of an operator which can be measured directly on a quantum processor (e.g. Pauli operator) takes a constant amount of measurements (i.e. shots) depending on the statistical precision that one wants to accomplish; however state preparation takes time proportional to the complexity of the circuit. This task then needs to be repeated as many times as the number of addends in the target operator decomposition from \cref{eq:vqe-operator-averaging}, which is why we require said amount to grow at most polynomially with the number of qubits. All in all, assuming polynomial circuit complexity with the number of ansatz parameters, the complexity of evaluating the objective function in VQE is $\order*{n^c N^q}$ for $\qty{c,q} \in {\mathbb{R}}^{+}$.

Worst case scenario, the optimizer will perform exhaustive search in the region of space spanned by our choice of ansatz parameters; which, expressing once again the exponential in base two, would result in a time complexity of the form:

\begin{gather}
  \textnormal{VQE}_\textnormal{exh} = \order{n^c N^q 2^{xn}} \, .
\end{gather}

Where $x$ is a complexity parameter, analogous to $s$ in \cref{eq:QSR-sample-parameter}, that represents the level of detail when searching the domain. On the other hand, assuming that the optimization process converges in polynomial time (e.g. convex optimization has many linear complexity algorithms \cite{NESTEROV.NEMIROVSKI:1994:interior-point-polynomial-algorithms-convex}, and the \emph{BFGS} method is known to possess quadratic time complexity \cite{NOCEDAL.WRIGHT:2006:numerical-optimization}), the overall complexity of VQE would be:

\begin{gather} \label{eq:VQE-complexity}
  \textnormal{VQE} = \order{n^c N^q n^p} \, .
\end{gather}

Where we can now see how the efficiency of VQE builds upon finding an economical parametrization ansatz possessing at most polynomial complexity with respect to the number of qubits (i.e. $n = \order*{N^t}$), so that the overall time complexity of VQE turns out polynomial as well. However, this is not an easy task; since the dimensionality of Hilbert space grows exponentially with the number of qubits.

Finally, the key insights to point out before moving on to QSR are: first, that the entire computation (i.e. each and every step of the optimization process) relies on the quantum processor, and therefore needs to make use of its resources heavily; and second, that the assumption of the optimization process converging in polynomial time is by no means expected to be true in most cases. In fact, the poly-time algorithms referenced above are generally not suitable for VQE, since objective functions are generally non-convex, and the random nature of quantum mechanics makes the evaluation of the expectation value function stochastic.


\subsection{Quantum Sampling Regression}\label{sub:qsr-complexity}

This algorithm must be divided into three distinct parts: sampling the target observable from the quantum processor, constructing the regression function, and performing the optimization. Each of which detached from one another.

Sampling the expectation value function takes time proportional to the complexity of the quantum circuit, the number of samples required, and the number of addends in the target's operator decomposition --- similar to VQE. For simplicity, we will use the upper bound from \cref{eq:QSR-algorithm-samples} to approximate the number of samples required by the algorithm. Assuming that the circuit has complexity polynomial with the number of ansatz parameters, this amounts to a time complexity $\order*{n^c N^q 2^{sn}}$. In turn, constructing the regression function also depends on the number of samples, having to solve a linear system of said size (i.e. cubic classical complexity): time complexity $\order*{2^{3sn}}$, and space complexity $\omega(2^{sn})$. Finally, the optimization process has to evaluate the regression at each step of the computation. Therefore, assuming polynomial complexity on the number of ansatz parameters for the optimizer we get time complexity $\order*{2^{sn} n^p}$; or, if we implemented the regression function in memory as to be able to evaluate it directly, $\Omega(n^p)$. All in all we have:

\begin{gather} \label{eq:QSR-complexity}
  \textnormal{QSR} = \order{n^c N^q 2^{sn} + 2^{3sn} + 2^{sn} n^p} \, .
\end{gather}

Because of the optimality of QSR (\cref{sec:algorithm-outline}), when comparing the quantum side of this result with that of exhaustive search in VQE, we can be sure that $x \ge s$ if using the same parametrization ansatz; otherwise the search would not be fine enough to capture all the necessary details of the objective function. This leads to QSR being asymptotically more efficient in terms of the amount of quantum resources it requires:

\begin{gather}
  \frac{\textnormal{VQE}_\textnormal{exh}}{\textnormal{QSR}} =
    \order*{2^{n\qty(x-s)}} \, .
\end{gather}

As for the classical side of the algorithm, it is dominated by solving the linear system of equations: $\order*{8^{sn}}$. For an ansatz where $n = \order*{N}$, this is similar to the complexity of solving the problem classically. If a better ansatz was to be found (e.g. $n=\order*{\log{N}}$), QSR would be more efficient in this regard as well --- as it harnesses the competence of the parametrization ansatz for classical computation. Finally, notice that, generally, $p_\textnormal{\tiny{QSR}} \leq p_\textnormal{\tiny{VQE}}$; since the optimization techniques accessible for a well-behaved deterministic regression function are not the same (i.e. better) as those for an stochastic objective function (\cref{sub:vqe-complexity}).

%% file: Content/04_Low-qubit-number-regime.tex

\section{Low qubit number regime}\label{sec:low-qubit-number-regime}

Because the algorithms at hand are thought to be valuable for limited quantum computational capabilities --- otherwise better algorithms such as \emph{phase estimation} would be available --- it becomes important to analyze their non-asymptotic behavior. According to the previous results, once we reach a high enough number of ansatz parameters (i.e. effective ansatz dimensions), it is expected that optimizing the objective function directly will be more time efficient than fully reproducing it; since, similar to exhaustive search, constructing the regression takes exponential time on top of that to optimize it. Also, for large bandwidths, and if a poly-time scheme is available in VQE, QSR might simply be less economical altogether. These questions have to be addressed quantitatively to find the practical limits of the method.

For this task, we are only going to consider the quantum side of the algorithms in order to see which one is more efficient in terms of the required amount of such resources. To do so, we will once again introduce the upper bound from \cref{eq:QSR-algorithm-samples}, and assume the most general monomial time complexity lower bound possible $(mn)^p$ for the VQE optimization process at low values of $n$. Since QSR optimization does not involve any quantum capabilities $p \equiv p_\textnormal{\tiny{VQE}}$. With this in mind, and from the quantum terms in \cref{eq:VQE-complexity,eq:QSR-complexity}, we can write:

\begin{gather} \label{eq:VQE-vs-QSR}
  \frac{\textnormal{VQE}}{\textnormal{QSR}} = \qty(m n 2^{-n/r})^{p} \, .
\end{gather}

Where $m$ has been included as part of VQE's complexity for generality, and we have defined $r \defeq p/s$. The motivation behind the latter is that, as we will see, most of the results extracted from this model do not depend on $p$ or $s$ alone, but rather on their quotient.

\begin{figure}[!tbp]
	\centering
	\begin{minipage}[c]{\linewidth}
		\centering
		\includegraphics[width=.75\linewidth]{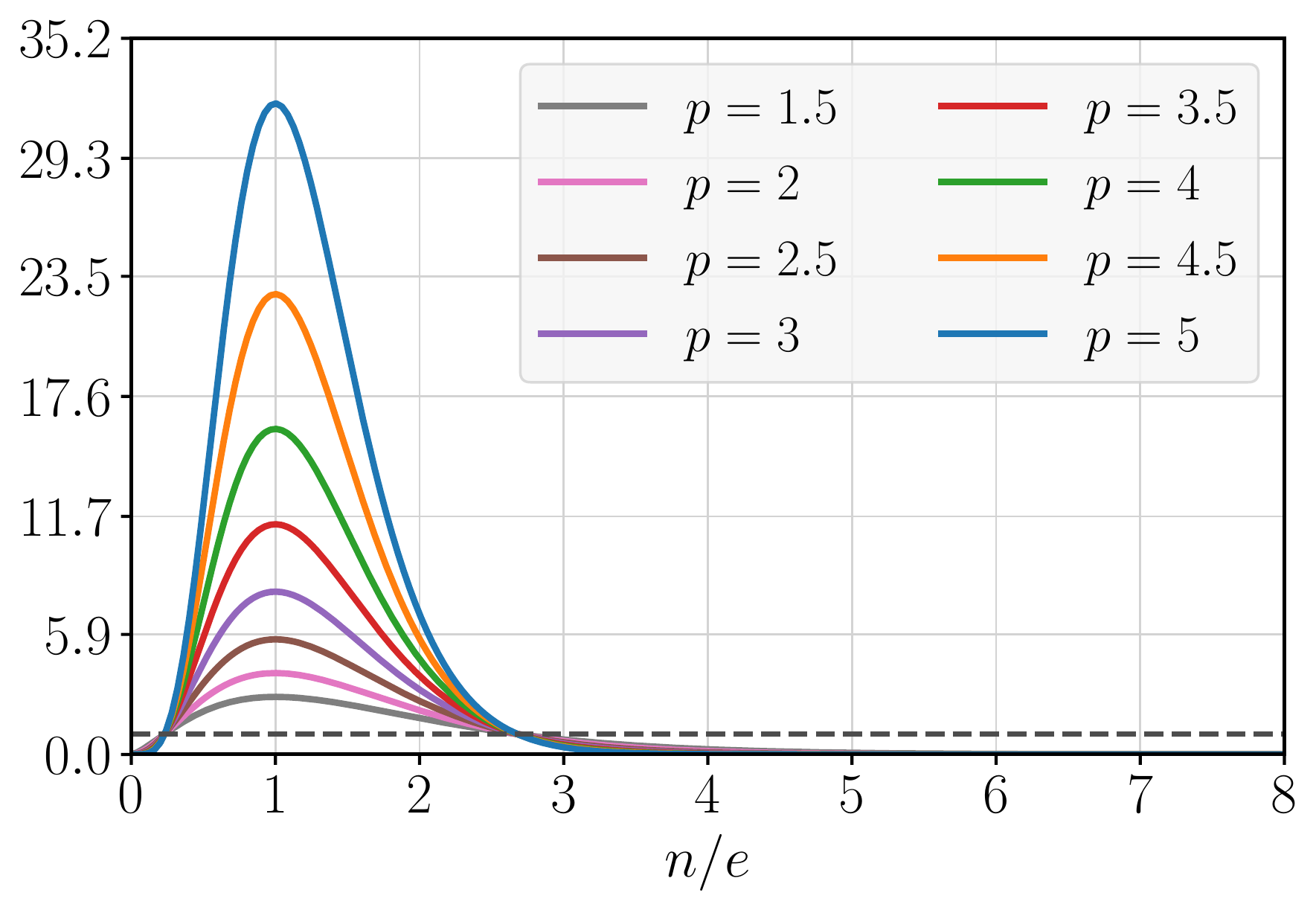}
	\end{minipage}
	\begin{minipage}[c]{\linewidth}
		\centering
		\includegraphics[width=.75\linewidth]{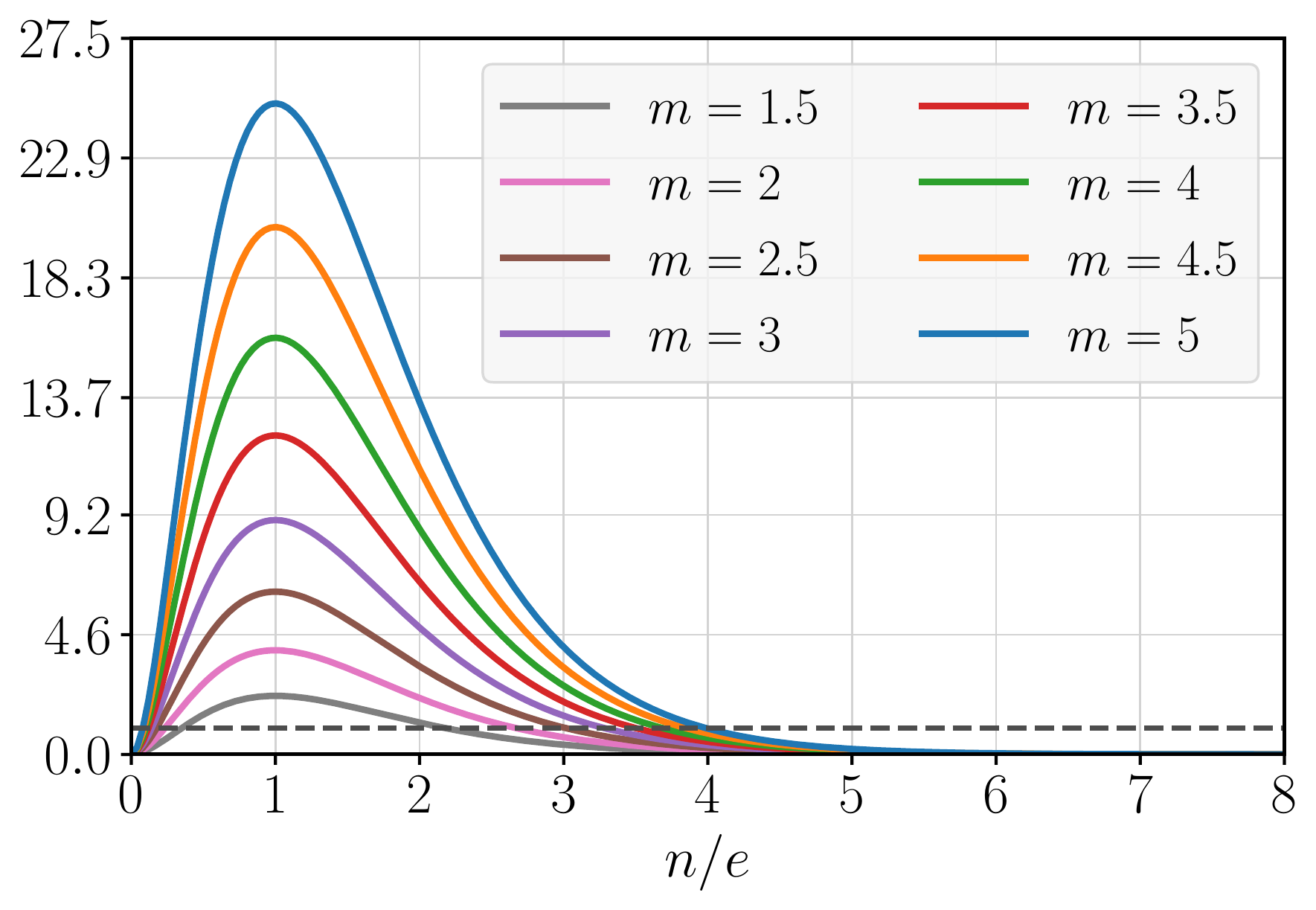}
	\end{minipage}
  \caption{Comparison between the amount of quantum resources required by VQE and QSR with respect to the number of parameters in the ansatz, according to \cref{eq:VQE-vs-QSR}, and assuming $r=e\ln{2}$. (Upper) Curves for different values of the VQE optimization complexity power $p$, fixing $m=2$. (Lower) curves for different values $m$, fixing $p=2$.}
  \label{fig:VQE-vs-QSR}
\end{figure}

This distribution will peak at low values of the number of ansatz parameters, as depicted in \cref{fig:VQE-vs-QSR}. For constant $r$, increasing $p$ while preserving $m$ will only make the peak taller (i.e. maintaining its same width); on the other hand, boosting $m$, will also widen it. Lastly, enlarging $r$ for constant $p$ and $m$ has the effect of scaling the graph: shifting its peak. We can now estimate to what number of ansatz parameters $n^{*} > 0$ the maximum will correspond, and its value in terms of $\qty{m,p,r}$:

\begin{gather}
  \dv{n} \qty(m n 2^{-n/r})^{p} =
    0 \qRa n^{*} = \frac{r}{\ln{2}} \label{eq:max-efficiency-parameters} \, , \\
  \eval{\frac{\textnormal{VQE}}{\textnormal{QSR}}}_{n=n^{*}} =
    \qty(\frac{mn^{*}}{e})^{p} \label{eq:max-efficiency} \, .
\end{gather}

From this we can now derive the critical conditions for QSR to be more efficient than VQE, at least for some range of the number of parameters in the ansatz.

\begin{gather}
  \eval{\frac{\textnormal{VQE}}{\textnormal{QSR}}}_{n=n^{*}} > 1 \qRa
    mn^{*} > e \, .
\end{gather}

To determine the range of ansatz parameters were QSR outperforms VQE, we need to find two values $n_{0,1} > 0$. For this, we can make use of the two real branches in the \emph{Lambert W function} or \emph{product logarithm} \cite{WOLFRAM-ALPHA:2001:generalized-lambert-function}:

\begin{gather}
  \eval{\frac{\textnormal{VQE}}{\textnormal{QSR}}}_{n=n_{0,1}} = 1 \qRa
    m n_{0,1} = 2^{n_{0,1}/r} \, , \\
  n_{0,1} = - n^{*} W_{0,-1}\qty(\frac{-1}{mn^{*}}) \, .
\end{gather}

Notice how \cref{eq:max-efficiency} is the only one which depends on $p$, meaning that this variable will merely affect the overall efficiency and not the number ansatz parameters for which QSR is more competent. As proof of correctness, we can see how $W_{0,-1}(x) < 0 \LRa x<0$, where $x \geq -1/e$; and for $x=-1/e$, we have that $mn^{*} = e$, $W_{0} = W_{-1}$, and $n_{0}=n_{1}=n^{*}$. It follows:

\begin{gather}
  \Delta \defeq n_1 - n_0 = n^{*}
    \qty[W_{0}\qty(\frac{-1}{mn^{*}}) - W_{-1}\qty(\frac{-1}{mn^{*}})] \, .
\end{gather}

Which, keeping in mind that, in practice, any valid number of ansatz parameters has to be a natural number (i.e. $n \in {\mathbb{N}}$), can now be used to find an estimate for the average efficiency gains in this region:

\begin{gather}
  E_\Delta \defeq
    \frac{1}{\floor{\Delta}}
    \sum_{n=\ceil{n_0}}^{\floor{n_1}} \frac{\textnormal{VQE}}{\textnormal{QSR}}
    \approx
    \frac{1}{\Delta}
    \int_{n_0}^{n_1} \frac{\textnormal{VQE}}{\textnormal{QSR}} \dd n \, .
\end{gather}

The above expression ranges from one to infinity (i.e. $E_\Delta \in \, ]1, \infty[ \,$), with typical values estimated as: $m \in \, ]0,10]$, $p \in [2,20]$, and $s \in [2,5]$. Furthermore, since QSR is an optimal instance of exhaustive search, we can define $a$ (based on $n_1$) as the threshold to overcome in order for quantum advantage to occur in VQE. In light of this, $n_0$ becomes irrelevant for characterizing practical use cases of the algorithm. With this in mind, we can simplify the above equations further, taking the positive efficiency $E \leq E_\Delta$ as:

\begin{gather}
  a \defeq \ceil*{- \frac{r}{\ln{2}} W_{-1}\qty(-\frac{\ln{2}}{mr})} \equiv
    \ceil{n_1} \label{eq:quantum-advantage-threshold} \, , \\
  E \defeq
    \frac{1}{a}
    \sum_{n=1}^{a} \frac{\textnormal{VQE}}{\textnormal{QSR}}
    \approx
    \frac{1}{a}
    \int_{1}^{a} \frac{\textnormal{VQE}}{\textnormal{QSR}} \dd n \, .
\end{gather}

The solution to this integral can ultimately be expressed in terms of the \emph{generalized upper incomplete gamma function} \cite{WOLFRAM-ALPHA:2001:generalized-incomplete-gamma-function}:

\begin{gather}
  E \approx \frac{1}{as\ln{2}} \qty(\frac{m}{s\ln{2}})^p
    \Gamma \qty(p+1, s\ln{2}, as\ln{2}) \label{eq:QSR-efficiency} \, .
\end{gather}

\begin{figure}[!btp]
  \centering
  \includegraphics[width=.75\linewidth]{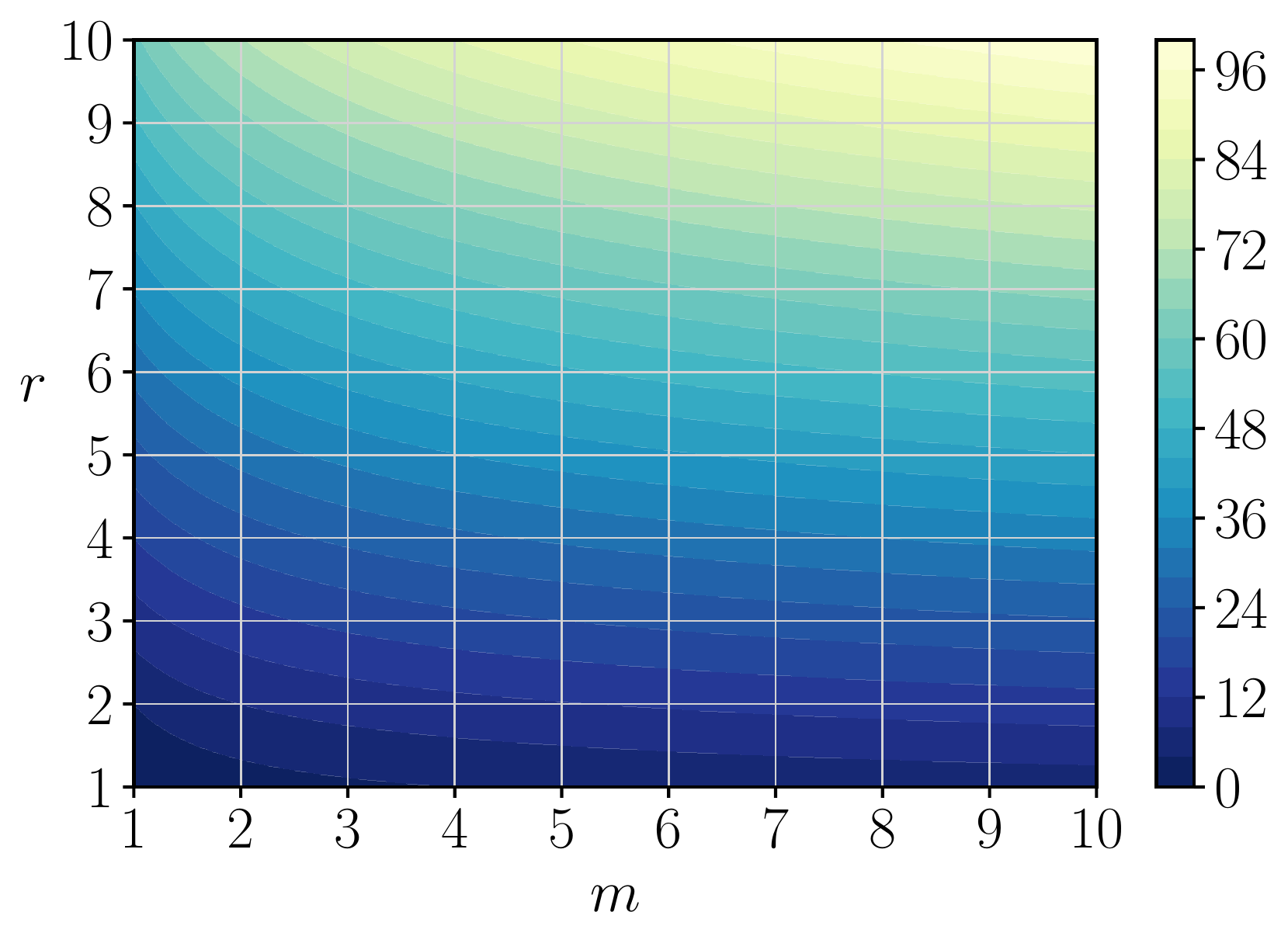}
  \caption{Color map of the quantum advantage threshold $a$ for different values of $m$ and $r$, as described by \cref{eq:quantum-advantage-threshold}.}
  \label{fig:quantum-advantage-threshold}
\end{figure}

\begin{figure}[!btp]
  \centering
  \includegraphics[width=.75\linewidth]{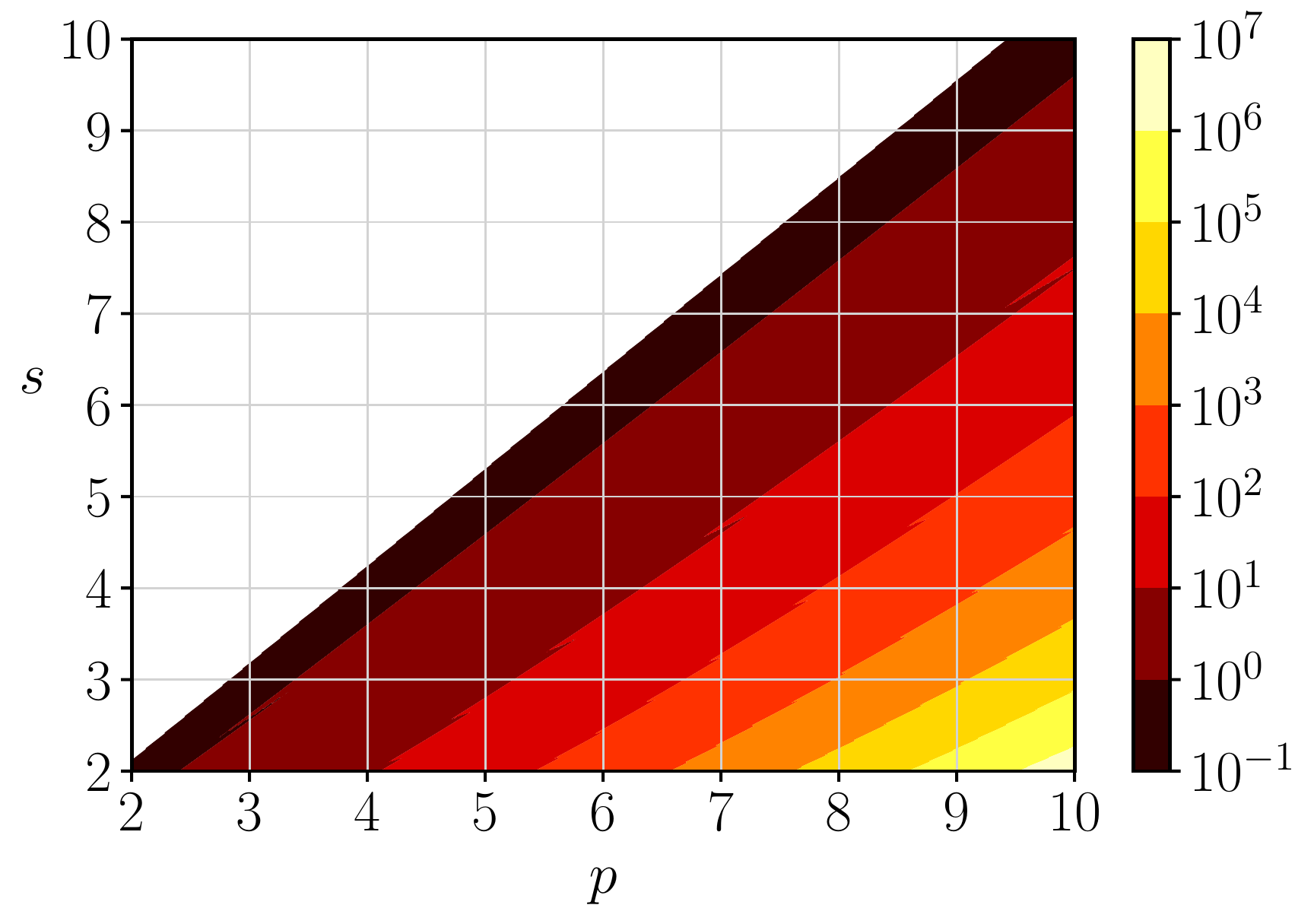}
  \caption{Color map showing the average efficiency gains, as described by the lower bound $E$ in \cref{eq:QSR-efficiency}, for different values of $p$ and $s$, and assuming $m=2$.}
  \label{fig:QSR-efficiency}
\end{figure}

\Cref{fig:quantum-advantage-threshold} depicts the quantum advantage threshold introduced in \cref{eq:quantum-advantage-threshold}, while \cref{fig:QSR-efficiency} shows different efficiencies stemming from \cref{eq:QSR-efficiency}. For the the values of $\qty{m,p,s}$ considered in the plots, it can be seen that, in particular cases, the threshold goes as high as a hundred ansatz parameters (i.e. one hundred effective dimensions in the ansatz), while the efficiency grows exponentially. This leads to massive average improvements in the total amount of samples required below said threshold, and it is therefore expected that QSR greatly outperforms VQE for certain problems --- especially those that can be tackled in state of the art quantum computers.

These results, although not exact, can be regarded as a simple, yet good analytical approximation, provided appropriate values for $\qty{m,p,s} \in {\mathbb{R}}^{+}$ (inasmuch as we only set a lower bound on the complexity of VQE's optimization process). As we mentioned in \cref{sec:algorithm-outline}, the value of $s$ can be obtained by analyzing the topology of the quantum circuit implementing certain parametrization ansatz. On the other hand, for a given problem, the values of $m$ and $p$ are retrieved by running VQE on an increasing amount of ansatz parameters $n$ and fitting a curve of the form $(mn)^p$ to lower bound the heuristic gathering the total number of samples required on each run. Along the same lines, detailed quantitative analysis could be easily conducted numerically, given a more precise description of the VQE optimizer for a problem of choice.

It is also worth mentioning that the $e$ and $\ln{2}$ factors come from our choice of basis for exponential complexity in the definition of $s$ (i.e. \cref{eq:QSR-sample-parameter}). The motivation behind said decision was to adopt the standard units of entropy in (quantum) information theory (i.e. bits) \cite{NIELSEN.CHUANG:2010:quantum-computation-quantum-information}, instead of natural units. In order to get rid of these terms in the equations, it is interesting to redefine:

\begin{gather}
  \tilde{n} \defeq \frac{n}{e} \qc \tilde{r} \defeq \frac{r}{e\ln{2}} \, .
\end{gather}

Leading to $\tilde{n}^{*} = \tilde{r}$; which, for \cref{fig:VQE-vs-QSR}, makes the scaling of the abscissa one to one with respect to the newly defined ratio, and the scaling of the ordinate $p$-polynomial: following the monomial used to lower bound VQE's complexity.

Finally, note that, even though we are assuming a low number of qubits (i.e. small $N$ implies limited $n$), this discussion is generally valid when the number of ansatz parameters is kept low --- even with large amounts of qubits. This could be achieved, if only in principle, with a good parametrization ansatz.

%% file: Content/05_Applications.tex

\section{Applications}\label{sec:applications}

While the VQE algorithm does not fully reconstruct the expectation value function, and could in principle require less sample points for solving the optimization (e.g. if the initial values were already close enough to the solution, or the number of ansatz parameters was above the threshold $a$), in many practical situations this will not be the case. The main explanation behind this is that classical optimizers usually need several evaluations of the objective function per iteration; which introduces a significant overhead, especially for relatively low values of the number of ansatz parameters.

Additionally, in most scenarios, it will be possible to control the frequency cutoff $s$ when designing the ansatz based state-preparation quantum circuit. Moreover, depending on the problem at hand, one might choose only a certain set of harmonics in the Fourier expansion if the rest are deemed irrelevant (e.g. choosing only even or odd components, or dropping the independent term); making QSR even more efficient by virtue of reducing the total number of samples $T$ necessary. If that was the case, \cref{eq:QSR-algorithm-samples} would have to be adapted.

This algorithm is therefore useful for reducing the amount of queries that need to be sent to a given quantum processor in order to carry out the optimization process. It imposes an optimal minimum number of samples to reproduce the quantum expectation value function classically; which can always be increased to attain higher precision (i.e. oversampling), or reduced to improve performance (i.e. undersampling). Furthermore, knowing all required samples from the beginning enables requesting them in as few queries as allowed by the backend (i.e. asking for several samples in just one connection); which could be accounted for in \cref{eq:VQE-vs-QSR} as an increase in $m$. This contrasts with VQE, where the optimization loop mandates new samples on each cycle; fundamentally impeding their compression in fewer requests.

This approach can prove specially useful for NISQ technologies, and generally for situations where not many quantum resources are available; reducing overheads and wait times to access quantum backends. At the same time, having a definite amount of evaluations to make can help us focus on improving the corresponding samples, mitigating noise further than what it is currently possible with regular VQE, and therefore championing improvements for results obtained through noisy machines.

Undersampling of the objective function implies imposing a frequency cutoff on its Fourier expansion, lower than that strictly necessary to fully reconstruct it. This non-exact strategy can still get decent approximate solutions, and will be especially useful in order to get rid of unnecessary small-wavelength oscillations leading to burdensome local minima, or if the Fourier coefficients tend to vanish for higher frequencies: a phenomenon well studied in Fourier theory \cite{KORNER:1998:fourier-analysis}. Notice that doing this implies reducing $s$ for constant $m$ and $p$ in \cref{eq:quantum-advantage-threshold,eq:QSR-efficiency}; improving the effectiveness of QSR with respect to VQE by explicitly violating \cref{eq:QSR-sample-parameter}, and in exchange for precision. The quality of such results could later be increased by using them as a better initialization state. In this way, we would regard QSR as a low-resolution algorithm to be iterated over (i.e. slashing the domain on every iteration), or even as a start-up supplement of VQE.

On top of that, QSR enables treating the quantum expectation value function as a black-box. The only information that needs to be known about it is the number of ansatz parameters it takes in, and the maximum frequency cutoff imposed by the circuit topology. No further knowledge about the actual operator that is being measured, or how it is being computed, is actually necessary --- maintaining a relatively high level of abstraction. With this approach, one can retrieve the objective function from the quantum realm and work with it (i.e. an approximation) classically, without relying on the usual exponentially-growing matrix formulation of quantum mechanics.

From the classical side of the computation, optimizing the regression function is far simpler than optimizing the randomized expectation values as exposed directly by the quantum processor. The reason for this being that, due to our flawed sampling system, the quantum computer returns a slightly different value each time we query it, even if evaluating for the same set of ansatz parameters. This implies that the objective function behaves stochastically when assessed directly from the quantum chipset, making optimization harder and less precise. As a matter of fact, QSR has shown to be helpful in practice for the optimization process to converge; since the random nature of quantum sampling causes all minima to slightly shift from one iteration to the next. This makes the entire VQE algorithm extremely sensitive to convergence conditions (i.e. forcing them to be more lax) and, by the same phenomenon, unwieldy for working with quantities such as (numerical) derivatives --- these again being essential for many optimization methods.

Lastly, quantum computing researchers often test their algorithms in specially engineered classical supercomputers often referred to as \emph{quantum simulators}: classical machines designed to emulate the behavior of quantum computers \cite{RAEDT.JIN.EA:2019:massively-parallel-quantum-computer}. Once they have debugged their code, they move on to running it in real quantum computers and dealing with the numerous problems that arise. The QSR algorithm, whose implementation can also be debugged on a simulator, could be used as a proxy to foresee and start tackling these concerns in a more controlled and isolated environment (i.e. the regression function). This would help analyze non-local effects (e.g. quantum gate error propagation), without having to deal with other setbacks such as the stochastic nature of function evaluation previously described, and without requiring the quantum processor to be available at all times --- since the Fourier coefficients could, in principle, be stored for later access. Altogether, this intermediate stage would imply a more restrained use of quantum resources; making the transition from simulators to real machines more sustainable.

All in all, we consider QSR to be especially useful for the current time and technology; particularly in scenarios where quantum resources are scarce and classical ones affordable, as well as those when convergence of the VQE algorithm becomes difficult or we need to deal with a smooth version of the objective function (e.g. for handling derivatives).

%% file: Content/06_Benchmarking.tex

\section{Benchmarking}\label{sec:benchmarking}

To benchmark this new technique, we will use \texttt{qiskit} \cite{ABRAHAM.AKHALWAYA.EA:2019:qiskit-open-source-framework-quantum} for reproducing the deuteron binding energy as described in \cite{DUMITRESCU.MCCASKEY.EA:2018:cloud-quantum-computing-atomic}. Using both VQE and the QSR algorithm with the minimum amount of samples approximated by \cref{eq:QSR-algorithm-samples}, we will expose some of the advantages of QSR on a problem of interest to the research community. Basing our benchmark on this work will also bring the opportunity to explore single and multiple ansatz parameter cases: paralleling our former explanation of the method. Through this exercise we will unveil how we are able to tell what the bandwidth is going to be by analyzing the quantum circuits implementing our parametrization; overall, what we will be fleshing out is a connection between the topology of the quantum circuit and that of the domain space in our parametrization as was introduced in previous sections. Although outside of the scope of the present work, this analysis of the topology of quantum circuits could potentially be formalized, for example, by means of the Directed Acyclic Graph (DAG) representation of quantum circuits \cite{CHILDS.SCHOUTE.EA:2019:circuit-transformations-quantum-architectures}. Furthermore, it could be conducted for standard ansatz implementations available in many quantum computing frameworks; pre-calculating their associated frequency bounds.

\begin{figure}[!tbp]
	\centering
	\begin{minipage}[c]{\linewidth}
		\centering
		\includegraphics[width=.75\linewidth]{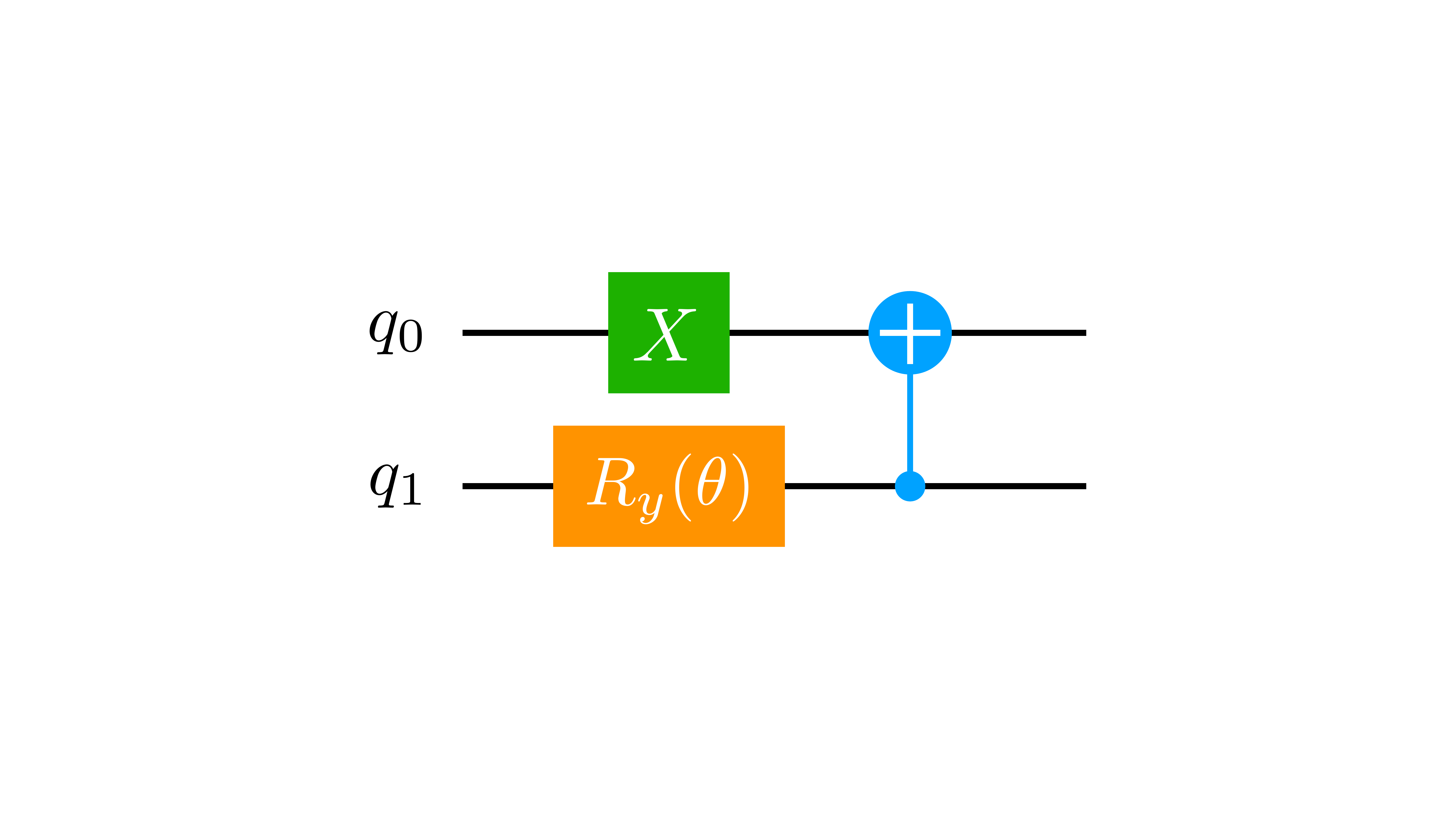}
	\end{minipage}
  \par\vspace{-3em}
	\begin{minipage}[c]{\linewidth}
		\centering
		\includegraphics[width=.75\linewidth]{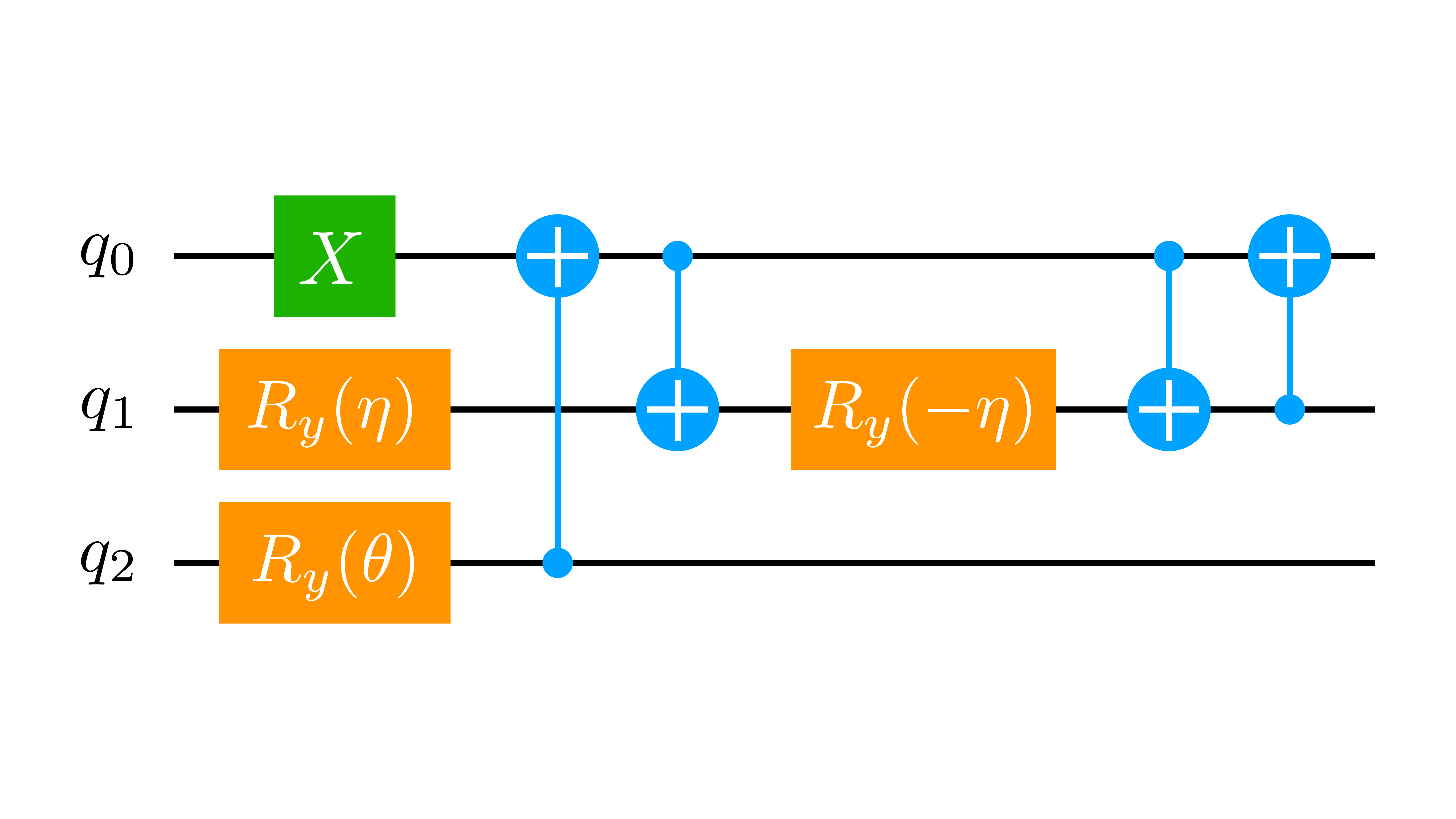}
	\end{minipage}
	\caption{Parametrization quantum circuits in reference \cite{DUMITRESCU.MCCASKEY.EA:2018:cloud-quantum-computing-atomic}, for one (upper) and two (lower) ansatz parameters.}
  \label{fig:deuteron-quantum-circuits}
\end{figure}

The purpose of these calculations is to approximate the deuteron binding energy on a quantum computer, by using a Hamiltonian from pionless effective field theory at leading order, and a problem-specific low-depth version of the unitary coupled-cluster ansatz. For said task, we will be dealing with two different observables: a two qubit Hamiltonian explored by a single parameter quantum circuit, and a three qubit Hamiltonian probed through a two parameters circuit. These are depicted in \cref{fig:deuteron-quantum-circuits}. In the first circuit, we notice that there is only one rotation and it does not interact with itself in any way. Therefore, we expect to find only the fundamental frequency, which can be easily verified from the original paper. The second circuit is slightly more complicated. This time, we find a rotation by an angle $\theta$ in qubit two, which propagates to qubit one (all the way through qubit zero) by means of two CNOT gates. In turn, qubit one starts off with yet another independent rotation, which then gets (partially) reversed after overlapping the previously described influence. For this reason, we expect the frequency bound of this second ansatz parameter $\eta$ to double (i.e. $S_{\eta}=2$). Meanwhile, the first rotation remains unaffected, so its respective frequency bound should stand equal to the fundamental frequency (i.e. $S_{\theta}=1$).

\begin{figure}[!tbp]
	\centering
	\begin{minipage}[c]{\linewidth}
		\centering
		\includegraphics[width=.75\linewidth]{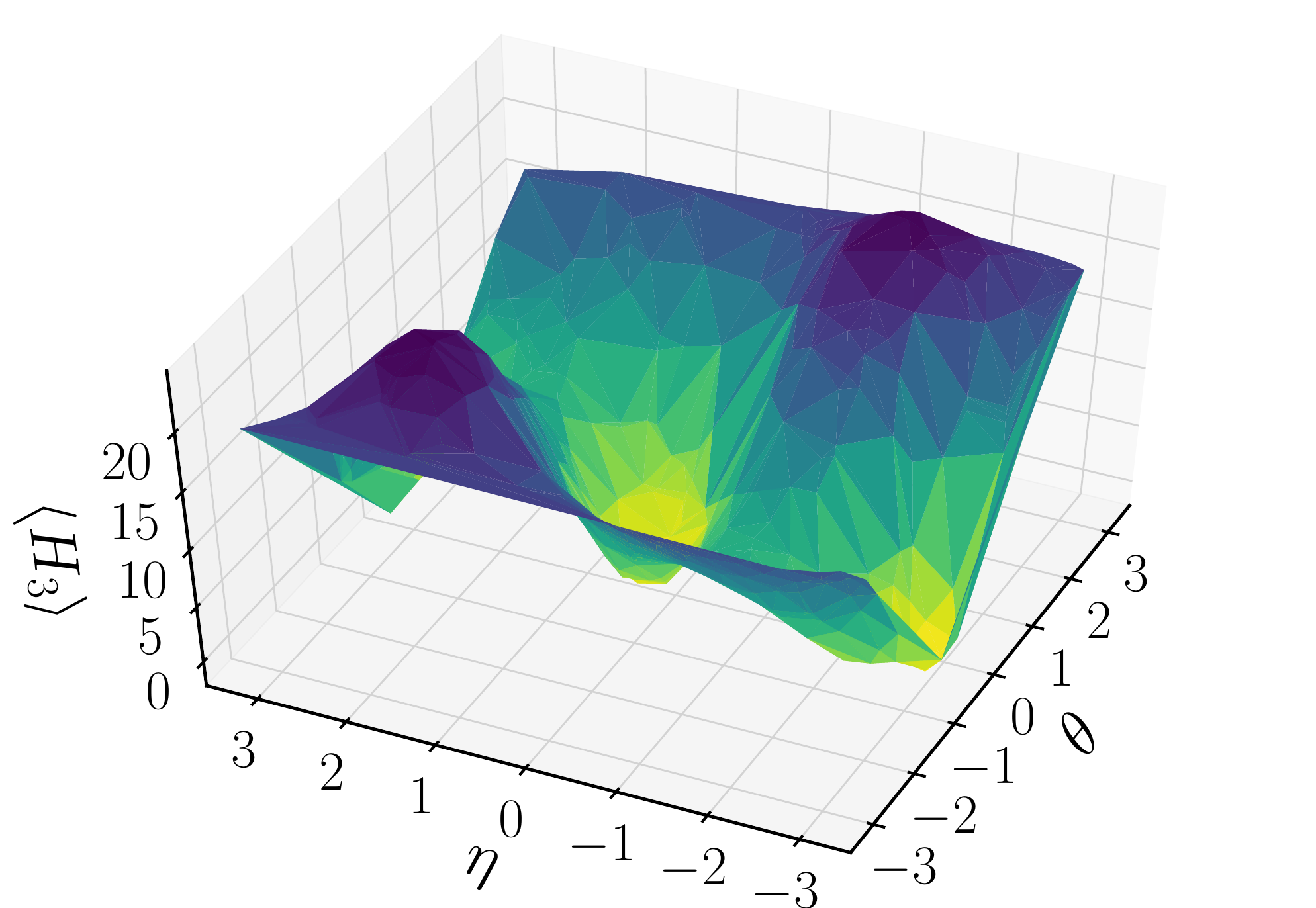}
	\end{minipage}
  \par
	\begin{minipage}[c]{\linewidth}
		\centering
		\includegraphics[width=.75\linewidth]{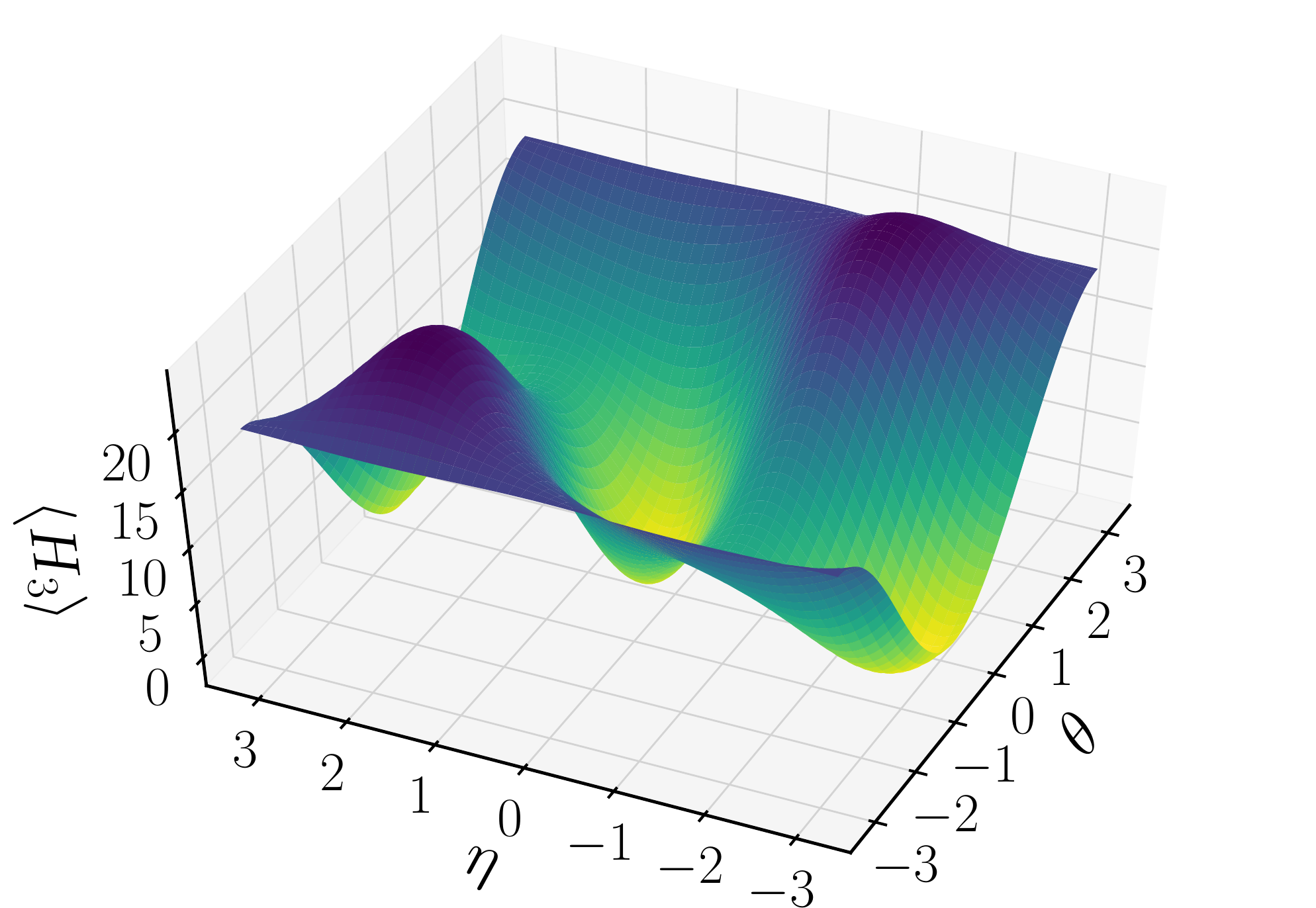}
	\end{minipage}
	\caption{Comparison between deuteron Hamiltonian expectation value as seen by the VQE and QSR algorithms, when reproducing the two ansatz parameters model in \cite{DUMITRESCU.MCCASKEY.EA:2018:cloud-quantum-computing-atomic}. (Upper) Triangulation of the expectation value function from raw samples. (Lower) Approximate function obtained through the Quantum Sampling Regression method with the minimum amount of samples approximated by \cref{eq:QSR-algorithm-samples}.}
  \label{fig:vqe-qsr-comparison}
\end{figure}

For the two ansatz parameters case, we have plotted a triangulation of the target function sampled directly from the quantum chip, against the regression function emerging from taking $S_q=S_{\text{max}}=2 ~\forall q$ in the multidimensional Fourier expansion (i.e. $T=25$), and evenly spaced samples: \cref{fig:vqe-qsr-comparison}. The results are compelling and, as portrayed in \cref{tab:vqe-qsr-comparison}, even manage to improve upon the original ones.

\begin{table}[!bp]
\caption{Comparison between results in \cite{DUMITRESCU.MCCASKEY.EA:2018:cloud-quantum-computing-atomic} reproduced using the VQE and QSR algorithms.\label{tab:vqe-qsr-comparison}}
	\begin{ruledtabular}
  \begin{tabular}{l l l l l}
  $n$ & Algorithm & Samples & Queries & Error \\
  \hline
  $1$ & VQE & $24$ & $24$ & $3.5\%$ \\
  $1$ & QSR & $3$ & $1$ & $1.0\%$
	\vspace{2pt} \\
  $2$ & VQE & $183$ & $183$ & $0.3\%$ \\
  $2$ & QSR & $25$ & $1$ & $0.2\%$ \\
  \end{tabular}
	\end{ruledtabular}
\end{table}

%% file: Content/07_Conclusions.tex

\section{Conclusions}\label{sec:conclusions}

The need to send HTTP requests over the internet for querying a quantum chipset introduces a pronounced bottleneck which greatly increases the time required to execute algorithms on actual hardware. Furthermore, the classical optimizers typically used in VQE are not particularly constrained in the amount of requests that they need to make, having to compute costly quantities (e.g. numerical derivatives) for intermediate steps: a concern that hardware providers are already trying to mitigate on their backends. This often trumps researchers looking to solve problems using said algorithm; which, at the same time, has established itself as the most promising near-term path to quantum advantage.

To address this issue, after a brief introduction to quantum variational eigensolvers (\cref{sec:introduction}), in this work we have proposed a new algorithm (QSR) based upon Fourier analysis of the expectation value function for a target quantum operator (\cref{sec:algorithm-outline}). The implemented strategy consists on building a regression function out of a minimum number of samples from the quantum processor, and optimizing it using classical means. Moreover, these samples will all be known beforehand and can therefore be requested in an even smaller number of queries to the quantum backend. We then studied our method's complexity as compared to classical simulations and VQE (\cref{sec:complexity-analysis}). Along the way, we also explained how QSR represents an optimal exhaustive search procedure on the given domain.

Since VQE is thought to be useful only for relatively scarce quantum resources (i.e. while phase estimation is prohibitive), we later introduced a mathematical model (\cref{sec:low-qubit-number-regime}) to compare its performance with that of our novel method in the low qubit number regime --- as opposed to the previous asymptotic analysis. From the same theoretical considerations in this survey, we inferred a quantum advantage threshold $a$, expressed in terms of our model's parameters. This threshold will therefore vary from problem to problem, depending on how complicated the optimization process turns out as depicted by complexity heuristics; and represents the amount of ansatz parameters below which QSR is more efficient than VQE for any given variational implementation.

Thereon, we outlined a number of different applications of this new technique (\cref{sec:applications}); highlighting its ability to get rid of local minima and boost performance by undersampling, as well as its possible role as a low-resolution supplement of VQE. On top of that, QSR can be used to wrap the quantum expectation value function as a classical black-box in order to avoid the exponential matrix formulation of quantum mechanics, while also capturing many of the effects intrinsic to experimental implementation. This can make optimization easier at the same time, improving convergence and precision. Researchers, may later find it valuable as a proxy to transition between simulators and real devices running VQE.

Finally, we presented the results of applying this new algorithm to reproduce the calculations of the deuteron binding energy from a previous paper (\cref{sec:benchmarking}). While doing so, we loosely discussed some of its implementation details and advantages, proving that QSR outperforms VQE in the same way predicted by our model.